\documentclass[manuscript,noblind]{geophysics}
\usepackage{soul}
\usepackage{amsmath}
\usepackage{amssymb}
\usepackage{algorithm}
\usepackage{multirow}
\usepackage{algpseudocode}
\usepackage{xcolor}
\usepackage{colortbl}
\usepackage{multirow}
\usepackage{ulem}
\usepackage{soul, color, xcolor}

\usepackage[colorlinks,
linkcolor=blue,
anchorcolor=blue,
citecolor=blue,backref]{hyperref}

\usepackage{lineno}
\begin{document}	
\title{SeisRDT: Latent Diffusion Model Based On Representation Learning For Seismic Data Interpolation And Reconstruction}



\author{Shuang Wang, Fei Deng, Peifan Jiang, Zezheng Ni and Bin Wang}


\maketitle

\begin{abstract}
Due to limitations such as geographic, physical, or economic factors, collected seismic data often have missing traces. Traditional seismic data reconstruction methods face the challenge of selecting numerous empirical parameters and struggle to handle large-scale continuous missing traces. With the advancement of deep learning, various diffusion models have demonstrated strong reconstruction capabilities. However, these UNet-based diffusion models require significant computational resources and struggle to learn the correlation between different traces in seismic data. To address the complex and irregular missing situations in seismic data, we propose a latent diffusion transformer utilizing representation learning for seismic data reconstruction. By employing a mask modeling scheme based on representation learning, the representation module uses the token sequence of known data to infer the token sequence of unknown data, enabling the reconstructed data from the diffusion model to have a more consistent data distribution and better correlation and accuracy with the known data. We propose the Representation Diffusion Transformer architecture, and a relative positional bias is added when calculating attention, enabling the diffusion model to achieve global modeling capability for seismic data. Using a pre-trained data compression model compresses the training and inference processes of the diffusion model into a latent space, which, compared to other diffusion model-based reconstruction methods, reduces computational and inference costs. Reconstruction experiments on field and synthetic datasets indicate that our method achieves higher reconstruction accuracy than existing methods and can handle various complex missing scenarios.
\end{abstract}
\section{Introduction}
Due to the limitations imposed by complex geological environments or economic factors, it is difficult to obtain complete seismic data, and collected seismic data usually exhibit missing traces \cite[]{chen2019interpolation}. This presents significant challenges for subsequent seismic data processing. Therefore, reconstructing complete seismic data is a crucial step in seismic data processing \cite[]{9703292,9768816}. Currently, methods for reconstructing complete seismic data are divided into two major categories: theory-driven and data-driven.

Traditional theory-driven methods mainly rely on the physical properties of the data and some assumed empirical parameters for interpolation. For example, the prediction filter method \cite[]{porsani1999seismic,gulunay2003seismic} transforms data into the f-x domain and uses the predictability of linear events in f-x domain to interpolate seismic data. Methods based on the wave equation require underground velocity as prior information for reconstructing seismic data \cite[]{fomel2003seismic,ronen1987wave}. However, obtaining an accurate subsurface velocity model is extremely challenging. Sparse constraint-based methods transform seismic data into the sparse domain and use compressed sampling theory for interpolation \cite[]{latif2016efficient,wang2010seismic}. Low-rank constraint methods based on compressed sensing are also employed \cite[]{zhang2019nonconvex,innocent2021robust}. However, these methods are not suitable for continuous missing traces. Most of the aforementioned traditional methods rely on the selection of some empirical parameters and are more suitable for randomly missing data \cite[]{niu2021seismic,wang2024seisfusion}, making it difficult to interpolate complex field data.

Data-driven methods based on deep learning primarily rely on the powerful learning capabilities of neural networks \cite[]{mousavi2022deep,mousavi2020earthquake} to interpolate seismic data \cite[]{jia2017can}. With the development of deep learning, there is increasing focus on the reconstruction performance of neural networks. \cite{wang2019deep} used ResNet \cite[]{he2016deep} to interpolate seismic data, demonstrating strong potential of neural networks for data interpolation. \cite{wang2020seismic} proposed using convolutional autoencoders to interpolate missing seismic data. \cite{zhang2020can} presented a reconstruction algorithm that combines deep learning with traditional methods, achieving good results in reconstructing small-scale missing data. \cite{chang2020seismic} introduced a dual-domain cGAN \cite[]{mirza2014conditional} for interpolating seismic data in the time-frequency domain. \cite{kaur2021seismic} introduced a self-learning seismic interpolation algorithm based on generative adversarial network (GAN). \cite{he2021seismic} trained a multi-stage UNet, achieving some success in interpolating low-amplitude missing components. \cite{yu2021attention} incorporated attention mechanisms and used a hybrid loss function, further enhancing the reconstruction capability of the UNet network.

As denoising diffusion probabilistic models (DDPM) have demonstrated powerful generative capabilities, their performance has been proven to surpass GAN networks \cite[]{creswell2018generative} and other point-to-point learning-based convolutional neural networks \cite[]{li2021survey} in the field of computer vision \cite[]{dhariwal2021diffusion}. The application of diffusion models to seismic data reconstruction has garnered attention. \cite{liu2024generative} used a conditional constrained diffusion model for seismic data interpolation. \cite{wei2023seismic} improved the sampling process of diffusion model, achieving good interpolation results. \cite{wang2024reconstructing} proposed a Classifier-Guided Diffusion Model for seismic data reconstruction, which improved reconstruction accuracy compared to convolutional methods. \cite{deng2024seismic} proposed a diffusion model that employs strong constraints to address the issue of unconditional random generation and improved the sampling process, achieving excellent reconstruction accuracy.

However, these state-of-the-art diffusion models for seismic data reconstruction perform the forward noise addition and reverse sampling processes in the data space \cite[]{yang2023diffusion}. This requires substantial computational resources and a significant amount of computation time \cite[]{wang2024patch,song2020denoising}. Therefore, inspired by the concept of stable diffusion \cite[]{rombach2022high}, using a trained data compression model to compress data into the latent space and then performing noise addition and sampling in latent space is a more reasonable choice \cite[]{vahdat2021score}. Secondly, their noise-matching network architectures are based on the inductive bias of UNet \cite[]{ronneberger2015u} architecture. Due to the limitations of convolutional kernel size, convolution operations have certain locality \cite[]{chen2021transunet,cao2022swin}, leading to the inability of UNet to perform global modeling of seismic data \cite[]{ma2024sit}. Furthermore, these UNet-based diffusion models learn the overall probability distribution of data, ignoring the correlations between different parts of the data \cite[]{gao2023masked}. Consequently, it is difficult to learn the relationships between different traces in seismic data, resulting in DDPM reconstructions where sampling is done from the latent space and then concatenated with existing data, rather than performing associative inference sampling from existing data. The Vision Transformer (ViT) \cite[]{dosovitskiy2020image} architecture demonstrates superior learning and generalization capabilities compared to traditional convolutional neural networks. Therefore, introducing the ViT architecture into diffusion models to replace the UNet architecture results in a more powerful generative model with global modeling capability—diffusion transformer (DiT) \cite[]{peebles2023scalable}.

To address these issues, we propose a latent diffusion transformer based on representation learning—seisRDT. First, we use a pre-trained data compression model to compress the data into latent space, followed by training and inference using a diffusion model in the latent space, which significantly reduces computational and inference costs. Secondly, we propose the Representation Diffusion Transformer architecture, granting diffusion models the ability for global modeling of seismic data. Compared to the convolutional UNet architecture, it demonstrates more powerful learning and generalization capabilities. Additionally, when calculating multi-head attention, we add a relative positional bias to each attention head, allowing the network to better focus on the positional information of the data. Finally, we propose a masked modeling \cite[]{he2022masked,cheng2021task} scheme based on representation learning \cite[]{devlin2018bert,radford2018improving}, where the representation module uses the token sequence of known data to infer the token sequence of unknown data, rather than simply sampling and concatenating. This leads to reconstructed data from the diffusion model having a more consistent data distribution, better correlation with known data, and higher accuracy. The contributions of this paper are as follows:

1. Propose a masked modeling scheme based on representation learning, enhancing diffusion models' ability to learn contextual information from seismic data. By utilizing the token sequence of known data through the representation module to infer and reconstruct the token sequence of unknown data, the reconstruction results achieve higher consistency and accuracy.

2. Propose the Representation Diffusion Transformer architecture, which includes the addition of a relative positional bias during the attention calculation, enabling diffusion models to gain global modeling capabilities. This allows for better learning of the probabilistic distribution of seismic data and provides stronger reconstruction capabilities.

3. By using a pre-trained data compression model, the training and inference processes of diffusion models are compressed into latent space. Compared to other diffusion model-based reconstruction methods, this approach significantly reduces computational and inference costs.
\section{Related Work} 
\subsection{Diffusion Model}
In the forward process, diffusion models perturb the data by gradually adding Gaussian noise \cite[]{sohl2015deep} through a fixed Markov chain \cite[]{kingma2021variational} until the data becomes isotropic Gaussian noise. Then, in the reverse process, a neural network is used to model the inversion of this process, progressively sampling from the noisy data to recover the clean data.

Given a data $x_0 \sim q(x_0)$, the forward process gradually transforms the data into a Gaussian noise $x_T \sim N(0, I)$ over $T$ time steps \cite[]{song2019generative} through a fixed Markov chain. A single step from $x_0$ to $x_t$ can be defined as:
 \begin{equation}
	q(x_t \mid x_{0})=\mathcal{N}(\sqrt{\overline{\alpha}_t}x_{0},(1-\overline{\alpha}_t)\mathbf{I})
\end{equation}
Using the reparameterization trick, $x_t$ can be expressed as:
 \begin{equation}
	x_t=\sqrt{\overline{\alpha}_t}x_{0}+\epsilon_t\sqrt{(1-\overline{\alpha}_t)},\epsilon_t \sim \mathcal{N}(0,\mathbf{I})
\end{equation}
where $\mathcal{N}$ indicates the Gaussian transition and $\mathbf{I}$ denotes the identity matrix, $\alpha_t =1-\beta_t, \overline{\alpha}_t =\prod_{s=0}^t \alpha_s $. The value of $\beta_t$ is obtained from a predefined variance table and typically linearly increases from 0.0001 to 0.002. $\epsilon_t$ is sampled from a Gaussian distribution.

The diffusion model employs a neural network to predict and reverse the encoding process. Its objective is to obtain the probability distribution $p_\theta(x_{t-1}| x_t)$ at time step $ t-1 $ as shown in Equation (3), where the mean $ \mu_\theta(x_t, t) $ and variance $ \beta_\theta(x_t, t) $ are derived from the neural network.
 \begin{equation}
	p_\theta(x_{t-1} \mid x_{t})=\mathcal{N}(\mu_\theta(x_t,t),\beta_\theta(x_t,t))
\end{equation}
According \cite{ho2020denoising}, $\mu_\theta(x_t,t)$ can be represented as:
\begin{equation}
	\mu_\theta(x_t,t)=\frac {1}{\sqrt{\alpha_t}}(x_t - \frac {\beta_t}{\sqrt{1-\overline{\alpha}_t}}\epsilon_\theta(x_t,t))
\end{equation}
$\beta_\theta(x_t,t)$ can be represented as:
\begin{equation}
	\beta_\theta(x_t,t)= \exp(\epsilon_\theta(x_t,t)\log\beta_t+(1-\epsilon_\theta(x_t,t))\log\tilde{\beta_t})
\end{equation}

Where $\tilde{\beta_t}=\frac {1-\overline{\alpha}_{t-1}} {1-\overline{\alpha}_{t}} \beta_t$.  $\epsilon_\theta(x_t,t)$ is the value predicted by the network.
During single-step training, the loss function of the network can be obtained by minimizing the variational lower bound of the negative log-likelihood.
 \begin{equation}
	L_{simple}=E_{t,x_0,\epsilon_t}[||\epsilon_t-\epsilon_\theta(x_t,t)||^2] 
\end{equation}
 Where $L_{simple}$ is the loss value used for network training and $\epsilon_t$ is given by Equation (2). After training, the diffusion model iteratively decodes the data starting from $x_T$ until the desired target data is obtained for the task.
\section{Method}
\renewcommand{\figdir}{Fig} 
\subsection{Latent Diffusion Model}
DDPM performs forward noise addition and reverse sampling processes in the original data space, resulting in significant computational time and resource requirements. Therefore, by introducing a data compression model to compress the data into latent space, training and sampling of diffusion models in the latent space can improve the model's generative quality while significantly reducing computational resource requirements.

We implement our data compression model based on vector quantized generative adversarial network (VQGAN) \cite[]{esser2021taming}, which is an improved model based on vector quantized variational autoencoder (VQVAE). It consists of an autoencoder training by perceptual loss and discriminative loss, as shown in Fig~\ref{fig:1}.
\plot{1}{width=\textwidth}
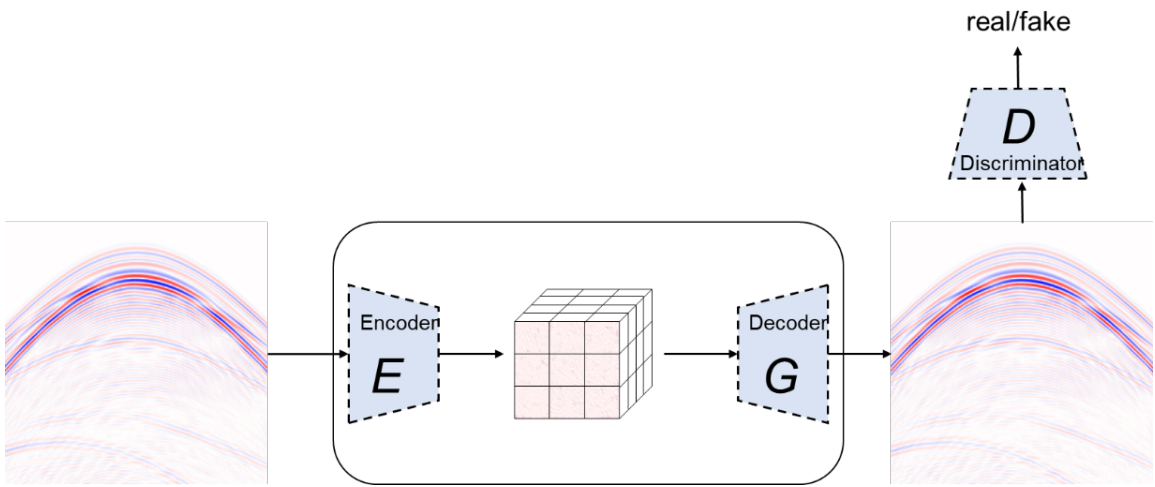
{Data compression model, seismic data are encoded into the latent space by E and decoded from the latent space by G to obtain the real data.}

Specifically, given a real seismic data \( x \), the encoder \( E \) encodes it into a representation in the latent space \( z = E(x) \), then the decoder \( G \) decodes data from the latent space, \( \widetilde{x} = G(z) = G(E(x)) \). We use perceptual loss to train this data compression model and introduce an adversarial training process using a discriminator \( D \). The training objective is represented as follows:
 \begin{equation}
	L=L_{rec}+L_{gan} 
\end{equation}
Where $L$ is the overall loss used for network training and \( L_{rec} \) represents the perceptual loss, which is composed of the $L2$ norm calculated by a pre-trained small-scale feature extraction network for extracting features from both the data \( x \) and the decoded data \( \widetilde{x} \). \( L_{gan} \) represents the discriminator loss, defined as follows, where $D$ denotes the discriminator:
 \begin{equation}
	L_{gan}=[logD(x)+log(1-D(\widetilde{x}))] 
\end{equation}
Once we have a powerful data compression model, we can compress the data into the latent space using the encoder \( E \). Then, we train the diffusion model in latent space and finally use decoder \( G \) to decode the \( z \) sampled by the diffusion model into real data. Training diffusion models in latent space is not significantly different from training in the real data space, as shown in Fig~\ref{fig:2}.
\plot{2}{width=\textwidth}
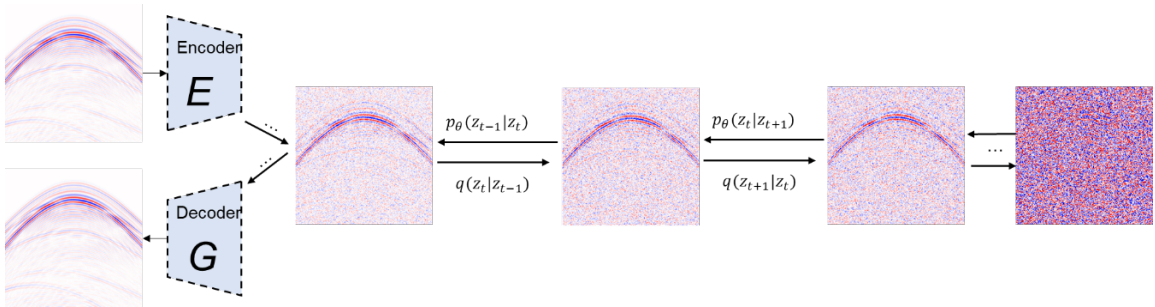
{data is encoded by the E of the data compression model into latent space, followed by the forward process q. After the reverse process p generates data, it is decoded by the decoder G into real data.}

The data compression model provides a compressed data \( z_0 \sim q(z_0) \), and the forward process adds noise to \( z_0 \) to get $z_t$.
 \begin{equation}
	z_t=\sqrt{\overline{\alpha}_t}z_{0}+\epsilon_t\sqrt{(1-\overline{\alpha}_t)},\epsilon_t \sim \mathcal{N}(0,\mathbf{I})
\end{equation}
The loss function of the diffusion model, focusing on the latent space, is re-expressed as:
 \begin{equation}
	L_{ldm}=E_{t,z_0,\epsilon_t}[||\epsilon_t-\epsilon_\theta(z_t,t)||^2] 
\end{equation}
Where $L_{ldm}$ is the loss value used for network training and $\epsilon_\theta(z_t,t)$ is the value predicted by the network.
\subsection{seisRDT}
In "Latent Diffusion Model" section, when modeling noise using neural networks in diffusion models, the commonly used architecture is UNet architecture with inductive bias. After introducing ViT, we obtained a powerful generative model, DiT, which possesses the ability to perform global modeling of seismic data. However, it learns the overall probability distribution of the data and ignores the relationships between the data. Therefore, we introduce representation learning to explicitly improve diffusion model's contextual learning ability. To avoid simply concatenating sampled data from the diffusion model with existing data during data reconstruction, we further propose a representation module. The representation module allows us to perform associative inference sampling from existing data, using known data to infer unknown data. This ensures that the reconstructed results have higher consistency and accuracy. The entire seisRDT reconstruction architecture is illustrated in Fig~\ref{fig:3}.
\plot{3}{width=\textwidth}
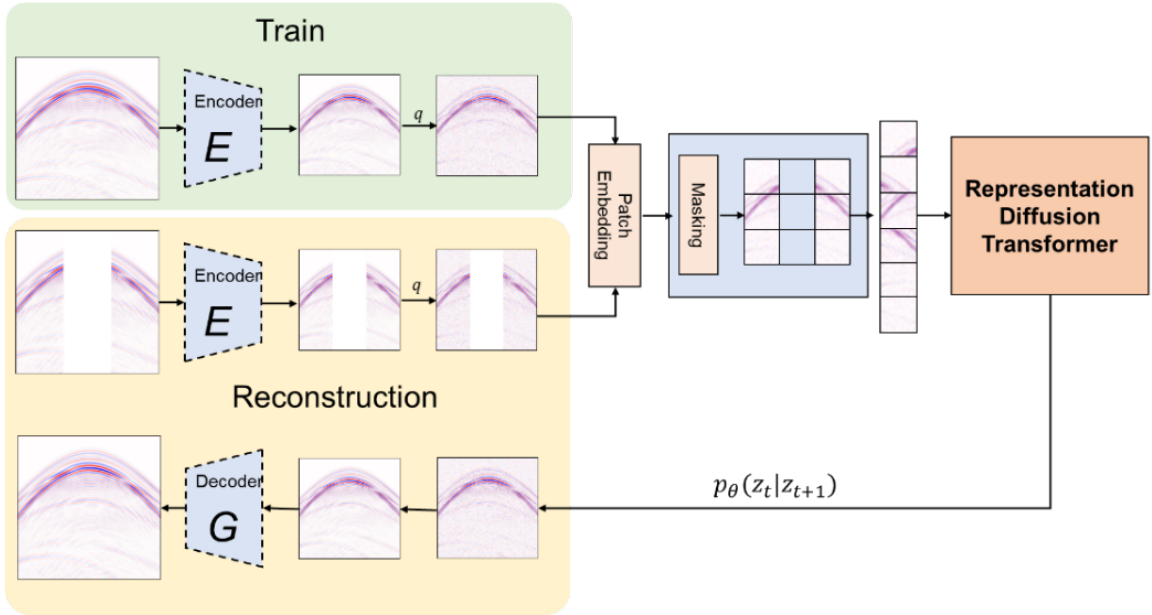
{The seisRDT architecture. Data compression model (left), the data undergoes compression, then Patch Embedding, followed by masking a portion. Subsequently, it undergoes inference by the Representation Diffusion Transformer (right).}

After data undergoes compression by the data compression model, the forward process of diffusion model is executed. Then, after the Patch Embedding operation, a portion of the data is masked. Subsequently, the Representation Diffusion Transformer (RDT) infers \( \epsilon_\theta(z_t, t) \). During training, complete data is input, and we actively perform masking operations to compel the model to learn the relationships between the data. During reconstruction, incomplete data is input, and existing data is fed into RDT for associative inference sampling. Next, we introduce the two key structures mentioned above: Patch Embedding and RDT.
\subsubsection{Patch Embedding}
After being compressed by the compression model, seismic data will obtain a latent representation \( z \) with a shape of \( I \times I \times C \). However, \( z \) is not a token sequence that can be accepted by the transformer. Therefore, Patch Embedding linearly embeds each patch in the input to transform \( z \) into a sequence of \( T \) tokens. The dimension of each token is \( d \), as shown in Fig~\ref{fig:4}.
\plot{4}{width=\textwidth}
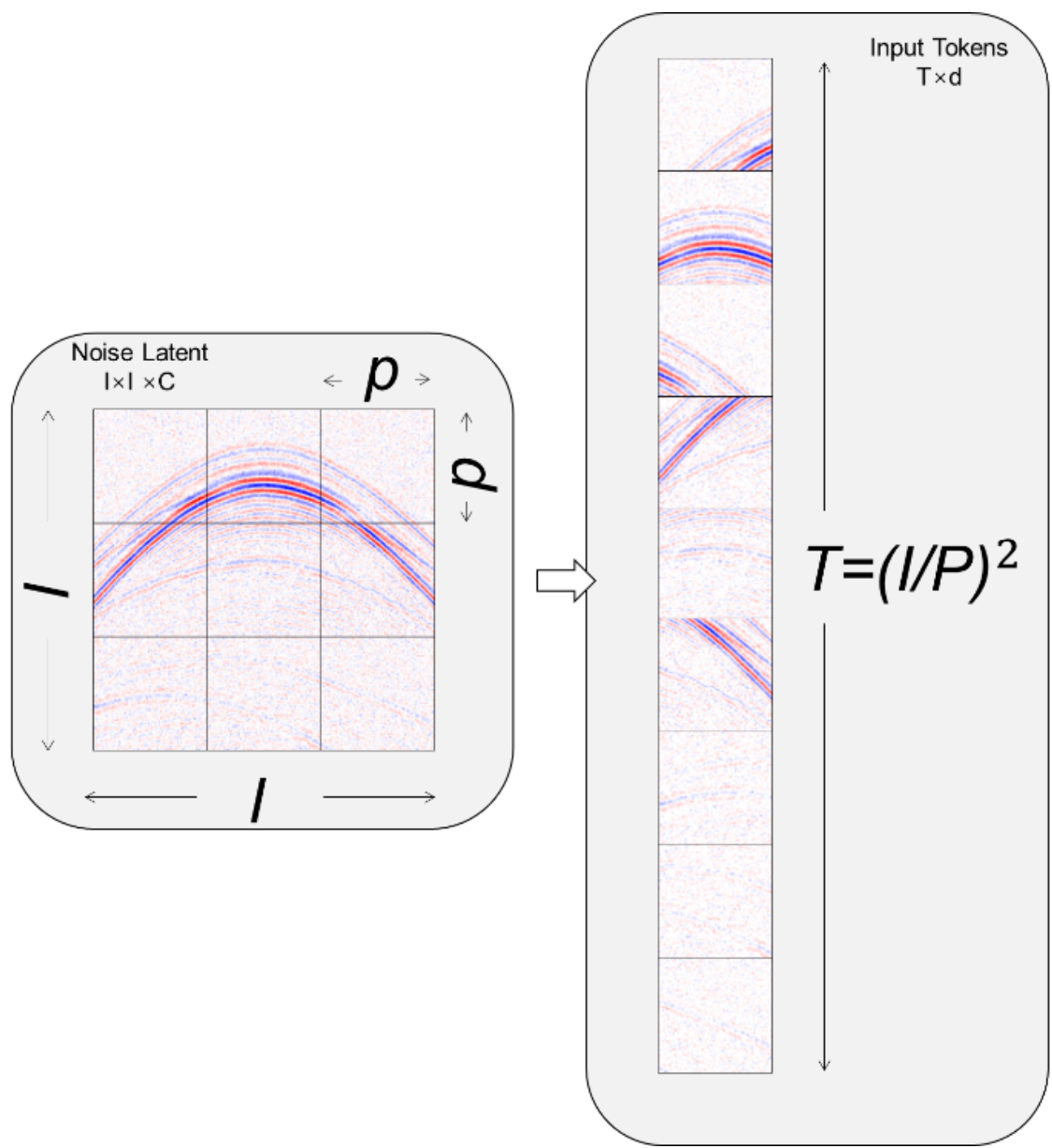
{Patch Embedding takes as input a tensor \( z \) of shape \( I \times I \times C \), with a given patch size \( p \times p \). It embeds this tensor into a sequence of length \( T = (I/p)^2 \), with a hidden dimension of \( d \).}

The length \( T \) of the token sequence obtained through Patch Embedding is determined by the hyperparameter \( P \). Halving \( P \) will quadruple the length of \( T \). Following the best practices of DiT, we set \( P \) to 2.
\subsubsection{Representation Diffusion Transformer}
The input of RDT is a token sequence obtained through Patch Embedding with a length of \( T \) and a dimensionality of \( d \). RDT consists of RDT Blocks and Representation Blocks, as shown in Fig~\ref{fig:5} (a). Firstly, we embed learnable global positional information into the token sequence, then encode it using N1 RDT Blocks. Next, the Representation Block is employed to infer unknown information using known information, and finally, the output is decoded using N2 RDT Blocks. The RDT Block follows the best practices of standard DiT, but to incorporate representation learning, we made some important modifications to it. The structure of the RDT Block is shown in Fig~\ref{fig:5} (b). When calculating multi-head attention after linear transformation, we added a relative position bias for each attention head:
 \begin{equation}
	Attention(Q,K,V)=Softmax(\frac{QK^T}{\sqrt{d_k}} + B_r)V
\end{equation}
\plot{5}{width=\textwidth}
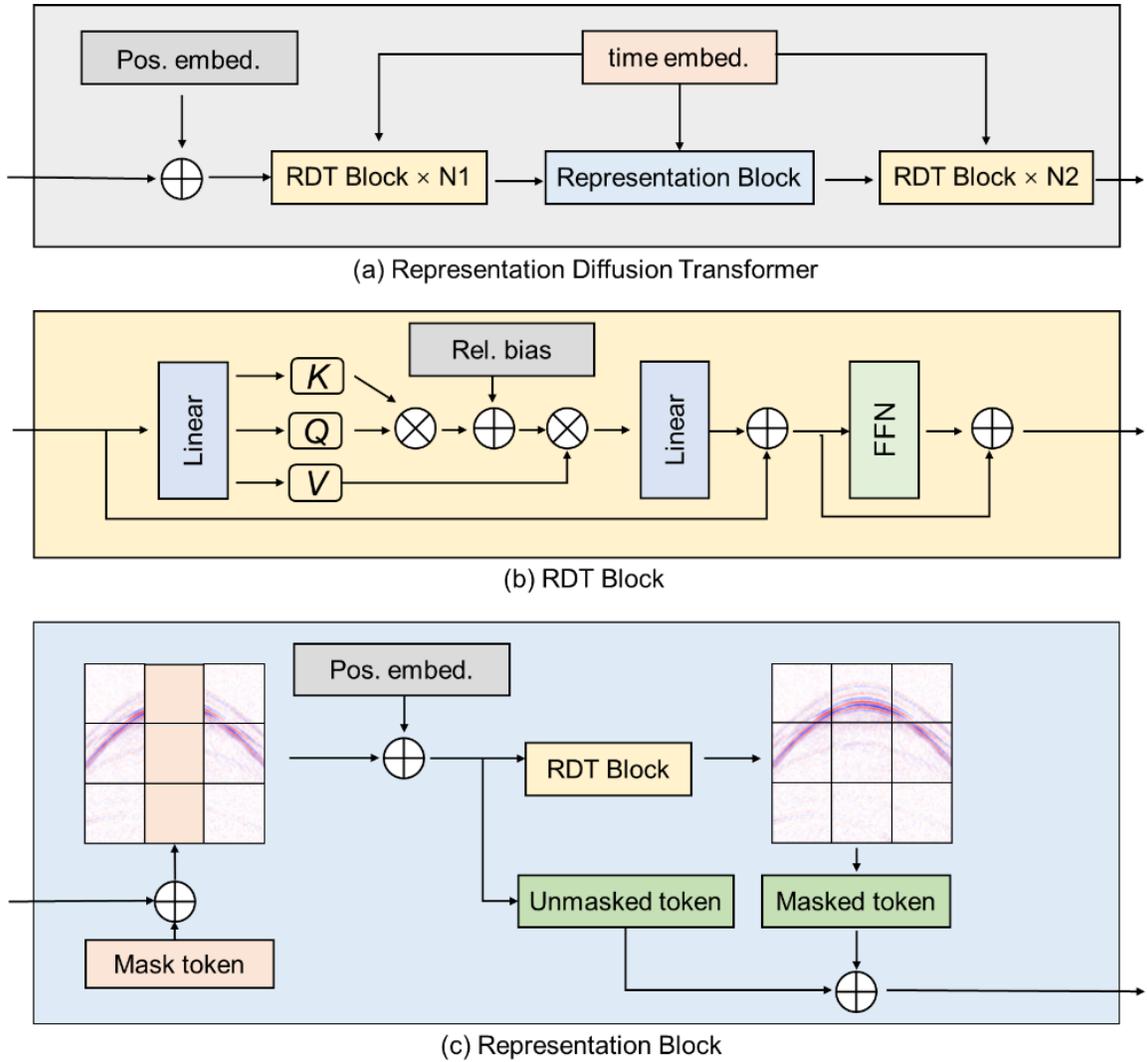
{The RDT architecture, from top to bottom: RDT structure, RDT Block structure, Representation Block structure.}
Where Q, K, and V represent the query, key, and value in the attention module, respectively, and \( d_k \) is their dimensionality. \( B_r \) is the relative positional bias \cite[]{liu2021swin} added for representation learning. \( B_r \) is selected from the bias table by the positional difference and is updated during training.Relative positional bias helps the model learn the relative relationships between different parts of the data. After the attention layer, linear transformation and residual connection are performed, followed by input to the feed-forward (FFN) module. The FFN module conducts multilayer perceptron (MLP) calculations, and then another residual connection is performed to obtain the output of the RDT block. The portion of the token sequence processed by N1 RDT Blocks during encoding is the part with masking operations applied, while during reconstruction, the token sequence encoded is the part of the data that is known. The Representation Block represents the output of the encoder as the complete token sequence expected by the decoder, which is then decoded by N2 RDT blocks.The structure of the Representation Block is shown in Fig~\ref{fig:5} (c). First, masked tokens are used to pad the incomplete token sequence, and position embedding is applied to enable the module to understand which tokens are known and which need to be represented as padded tokens. Next, a basic RDT block is used to process and predict the tokens that need to be padded. Finally, a shortcut connection is used to combine the predicted token sequence with the known token sequence that has been embedded with position information. In summary, for masked tokens, we use the predictions from the RDT block. For unmasked tokens, we continue to use the existing token sequence.
\section{Experiments}
\renewcommand{\figdir}{Fig} 
\subsection{Evaluation Metrics}
To quantitatively evaluate the quality of the reconstruction results, we selected three commonly used metrics: mean squared error (MSE), signal-to-noise ratio (SNR), and structural similarity index (SSIM). MSE measures the error between the reconstruction results and the ground truth, and its calculation formula is as follows:
 \begin{equation}
	MSE=\frac {1}{n}\sum_{i=1}^{n}(x_{i}^{r}-x_{i}^{t})^2
\end{equation}
Where \( x_i^r \) represents the data reconstructed by the network, \( x_i^t \) denotes the ground truth. The closer the value of MSE is to 0, the closer the reconstruction result is to the ground truth. SNR measures the quality of the reconstruction result, and its calculation formula is as follows:
 \begin{equation}
	{SNR=10log_{10}\frac{\parallel x_t \parallel_F^2 }{\parallel x_t-x_r \parallel_F^2}}
\end{equation}
Where \( x_r \) represents the data reconstructed by the network, \( x_t \) denotes the ground truth, and \( {||\ ||}_F \) represents the Frobenius norm. A larger SNR value indicates higher quality of the reconstruction result. SSIM measures the structural consistency of the reconstruction result, and its calculation formula is as follows:
 \begin{equation}
	SSIM=\frac{(2\mu_r\mu_t+c_1)(2\sigma _{rt}+c_2)}{(\mu_r^2\mu_t^2+c_1)(\sigma _r^2+\sigma _t^2+c_2)} 
\end{equation}
Where \( u_r \) is the mean of the reconstructed result, \( u_t \) is the mean of the ground truth, \( \sigma_{rt} \) is the covariance between the reconstructed result and the ground truth, \( \sigma_r \) is the variance of the reconstructed result, \( \sigma_t \) is the variance of the ground truth, \( c_1 \) and \( c_2 \) are two constants introduced to avoid numerical instability. A higher SSIM value closer to 1 indicates that the reconstruction result is closer to the ground truth.
\subsection{Train}
We conducted experiments on the publicly available SEG C3 synthetic dataset and the Mobil Avo Viking Graben Line 12 field dataset. The SEG C3 synthetic dataset contains a total of 45 shot gathers, each with a 201×201 receiver grid. The sampling rate is 8ms, with 625 samples per trace. We randomly cropped 2000 patches, each of size 256×256, with 1600 used for training and 400 for testing. The Mobil Avo Viking Graben Line 12 field dataset has a receiver grid of size 1001×120. The sampling rate is 4ms, with 1500 samples per trace. We randomly cropped 2000 patches, each of size 256×256, with 1600 used for training and 400 for testing. We first trained the data compression model, training it as an encoder E to compress data $x$ of size 256×256 into a latent representation $z$ of size 32×32, and a decoder G to recover $x$ from the latent representation $z$ with high quality. The compression ratio directly affects the computational cost and the quality of the generated data. When the compression ratio is small, such as 1 or 2, the size of the latent space data remains large, resulting in higher computational costs. Conversely, an overly large compression ratio, such as 16 or 32, can lead to information loss, making it difficult for the diffusion model to perform high-precision reconstruction. Therefore, selecting an appropriate compression ratio is crucial. For the trade-off ratios of 4 and 8, choosing 8 is a better option. This ratio compresses the 256×256 data to 32×32, significantly reducing the computational cost while maintaining relatively high reconstruction accuracy. Furthermore, setting the compression ratio to 8 has been proven to be an optimal choice in some two-stage generative models \cite[]{rombach2022high}. For training the dataset compression model, we set the batch size to 8, used AdamW as the optimizer, and set the learning rate to 0.0001. Training was conducted for 200 epochs on a single NVIDIA A6000 GPU. Then we used the compressed latent representation $z$ to train the Representation Diffusion Transformer, with p set to 2 during Patch Embedding. Clearly and unquestionably, this approach saves a significant amount of computational and inference costs compared to directly training the diffusion model using the original data $x$ of size 256×256. We trained for 200,000 steps on an Nvidia A6000 GPU. We set N1 to 24, N2 to 8, and the batch size to 16, using AdamW as the optimizer with a learning rate of 0.0001. We employed two masking strategies: random masking and equidistant masking. For random masking, the masking rate was set between 0.2 and 0.3, which helps the model learn the relationships between different parts of the data. For equidistant masking, the masking rate was set between 0.2 and 0.7 with an interval of 16, meaning that if the current token is masked, every 16th token from that position will also be masked. This simulates the missing data scenario in seismic data.

We selected three of the most advanced reconstruction methods utilizing diffusion models for comparative testing: SeisDDIMR \cite[]{wei2023seismic}, DPM-based interpolation method (DPM) \cite[]{liu2024generative}, and Conditional Constraint Diffusion Model (CCDM) \cite[]{deng2024seismic}. SeisDDIMR modifies the sampling process of the diffusion model by concatenating known and sampled data and then performing iterative sampling for data reconstruction.DPM adds conditional constraints to an unconditional diffusion model by placing known and sampled data into separate channels, with the network using a two-channel input. CCDM not only adds conditional constraints but also alters the diffusion model's sampling process. In CCDM, the conditional constraints are handled by a separate network, and the sampled data is processed by another network. The results from these two networks are then combined to estimate the noise. During sampling, the data is concatenated with known and sampled data before iterative sampling. We used the exact same dataset and set their hyperparameters according to the descriptions in their respective papers. Similarly, we trained them for 200,000 steps on an Nvidia A6000 GPU before conducting the comparisons. Additionally, we included a UNet based method, Anet \cite[]{yu2021attention}, and a traditional reconstruction method, POCS. Anet modifies the network’s loss function by incorporating hybrid loss and introduces attention into UNet, and was trained on the same dataset, optimized according to the settings described in the paper.
\subsection{Test}
\subsubsection{Synthetic Dataset}
To fully validate the effectiveness of the proposed method, we designed two missing data scenarios: random missing and continuous missing. 

Random missing:
We set two missing ratios, 40\% and 70\%, by setting traces to 0 to simulate missing traces. Fig~\ref{fig:6} illustrates the reconstruction results of different methods under these two missing ratios.
\plot{6}{width=1.1\textwidth}
{(a) Ground truth,(b) 40\% random missing, (b-1) to (b-6) are the reconstruction results of POCS, Anet, DPM, SeisDDIMR, CCDM, and SeisRDT, respectively. (b-7) to (b-12) are the residuals of each reconstruction result relative to the ground truth.(c) 70\% random missing, (c-1) to (c-6) are the reconstruction results of POCS, Anet, DPM, SeisDDIMR, CCDM, and SeisRDT, respectively. (c-7) to (c-12) are the residuals of each reconstruction result relative to the ground truth.}

It can be observed that the POCS method has relatively poor accuracy, when reconstructing with 70\% random missing data, the reconstruction of a slightly larger gap even failed. Anet, compared to the other diffusion model based methods, shows slightly lower reconstruction accuracy. The four diffusion model methods achieved satisfactory data reconstruction. However, according to the residual plots, our method achieved the highest reconstruction accuracy, with the smallest deviation from the ground truth. For quantitative evaluation, we listed three evaluation metrics for the six methods under random missing conditions in Table~\ref{tbl:1}. Our method maintains a certain level of superiority in all metrics, indicating that seisRDT has better reconstruction accuracy.
\tabl{1}{Comparison of six reconstruction methods under random missing on synthetic dataset. The upper section compares 40\% randomly missing traces, while the lower section compares 70\% randomly missing traces.}{
\begin{center}
\linespread{1.25} \selectfont
\begin{tabular}{ccccc}
	\hline
	&                 & MSE       & SNR     & SSIM   \\ \hline	
	& POCS  & 3.1473e-4    & 41.0411 & 0.9539 \\
	& Anet & 1.9344e-4 & 43.1549 & 0.9718 \\  
	& DPM            & 6.2493e-5    & 47.0622 & 0.9955 \\
	& SeisDDIMR & 6.4499e-5 & 47.9250 & 0.9954 \\
	& CCDM         & 6.3311e-5 & 47.0057 & 0.9956 \\
	& SeisRDT        & 2.9294e-5 & 51.3527 & 0.9971 \\ \hline
	& POCS            & 0.0131    & 24.8430 & 0.8356 \\
	& Anet 	& 8.6976e-4 & 37.1058 & 0.9356 \\  
	& DPM            & 3.2574e-4    & 40.8918  & 0.9823 \\
	& SeisDDIMR & 2.4400e-4    & 41.1465 & 0.9849 \\
	& CCDM         & 1.7321e-4 & 43.6348 & 0.9862 \\
	& SeisRDT        & 4.9436e-5 & 49.0801 & 0.9946 \\ \hline
\end{tabular}
\end{center}
}

Continuous missing:
We set two scenarios of continuous missing, with 40 and 70 consecutive traces set to 0 to simulate missing traces. Fig~\ref{fig:7} shows the reconstruction results of different methods.
\plot{7}{width=1.1\textwidth}
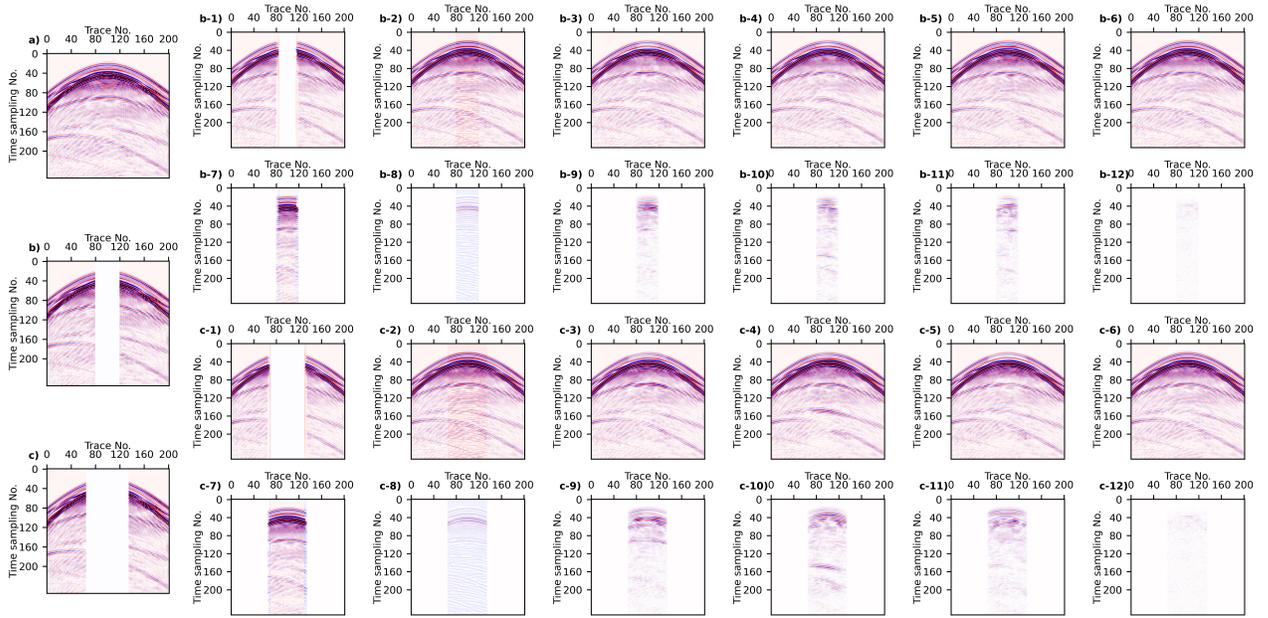
{(a) Ground truth,(b) 40 continuous missing traces, (b-1) to (b-6) are the reconstruction results of POCS, Anet, DPM, SeisDDIMR, CCDM, and SeisRDT, respectively. (b-7) to (b-12) are the residuals of each reconstruction result relative to the ground truth.(c) 70 continuous missing traces, (c-1) to (c-6) are the reconstruction results of POCS, Anet, DPM, SeisDDIMR, CCDM, and SeisRDT, respectively. (c-7) to (c-12) are the residuals of each reconstruction result relative to the ground truth.}

Under continuous missing data, the traditional method failed in reconstruction, and Anet exhibited some inconsistencies. However, all four diffusion model methods achieved satisfactory data reconstruction. However, according to the residual plots, our method demonstrated the highest reconstruction accuracy with the smallest deviation from the ground truth. For quantitative assessment, we listed three evaluation metrics for the six methods under continuous missing scenarios in Table~\ref{tbl:2}. The results indicate that our method exhibits the highest reconstruction accuracy.
\tabl{2}{Comparison of six reconstruction methods under continuous missing on synthetic dataset. The upper section compares 40 continuous missing traces, while the lower section compares 70 continuous missing traces.}{
	\begin{center}
		\linespread{1.25} \selectfont
		\begin{tabular}{ccccc}
			\hline
			&                 & MSE       & SNR     & SSIM   \\ \hline
			& POCS  & 0.1152    & 15.4034 & 0.8499 \\
			& Anet & 4.4966e-3 & 39.7281 & 0.9661 \\ 
			 & DPM            & 2.6864e-4    & 41.7288 & 0.9890 \\
			& SeisDDIMR & 1.6898e-4 & 43.7438 & 0.9885 \\
			& CCDM         & 1.6179e-4 & 43.9309 & 0.9905 \\
			& SeisRDT        & 5.9294e-5 & 47.3527 & 0.9986 \\ \hline
			& POCS  & 0.2415    & 12.1908 & 0.7281 \\
			& Anet & 8.4635e-3 & 36.3665 & 0.9270 \\  
			& DPM            & 4.5606e-4    & 39.4303  & 0.9709 \\
			& SeisDDIMR & 4.3438e-4    & 39.6418 & 0.9729 \\
			& CCDM         & 3.2338e-4 & 40.9233 & 0.9791 \\
			& SeisRDT        & 1.0436e-4 & 44.1472 & 0.9906 \\ \hline
		\end{tabular}
	\end{center}
}

To facilitate a better comparison of the six methods, we plotted the f-k spectra under the scenario of 70 continuous missing traces, as shown in Fig~\ref{fig:8}. It can be observed that our method closely resembles the ground truth, with minimal artifacts. The traditional method performs the worst, while Anet and the other three diffusion model based methods exhibit some spatial aliasing. This indicates the effectiveness of the SeisRDT in reconstruction seismic data with missing traces.
\plot{8}{width=\textwidth}
{f-k spectra. (a) 70 continuous missing traces, (b) reconstruction result of POCS, (c) reconstruction result of Anet, (d) reconstruction result of DPM, (e) reconstruction result of SeisDDIMR, (f) reconstruction result of CCDM, (g) reconstruction result of SeisRDT, (h) ground truth.}

To better illustrate the amplitude preservation of the SeisRDT, Fig~\ref{fig:9} shows the reconstruction of the 100th trace under the scenario of 70 continuous missing traces for the six different methods. It can be observed that the SeisRDT reconstructs local detailed information effectively, with the reconstructed data closely resembling the original seismic record.
\plot{9}{width=\textwidth}
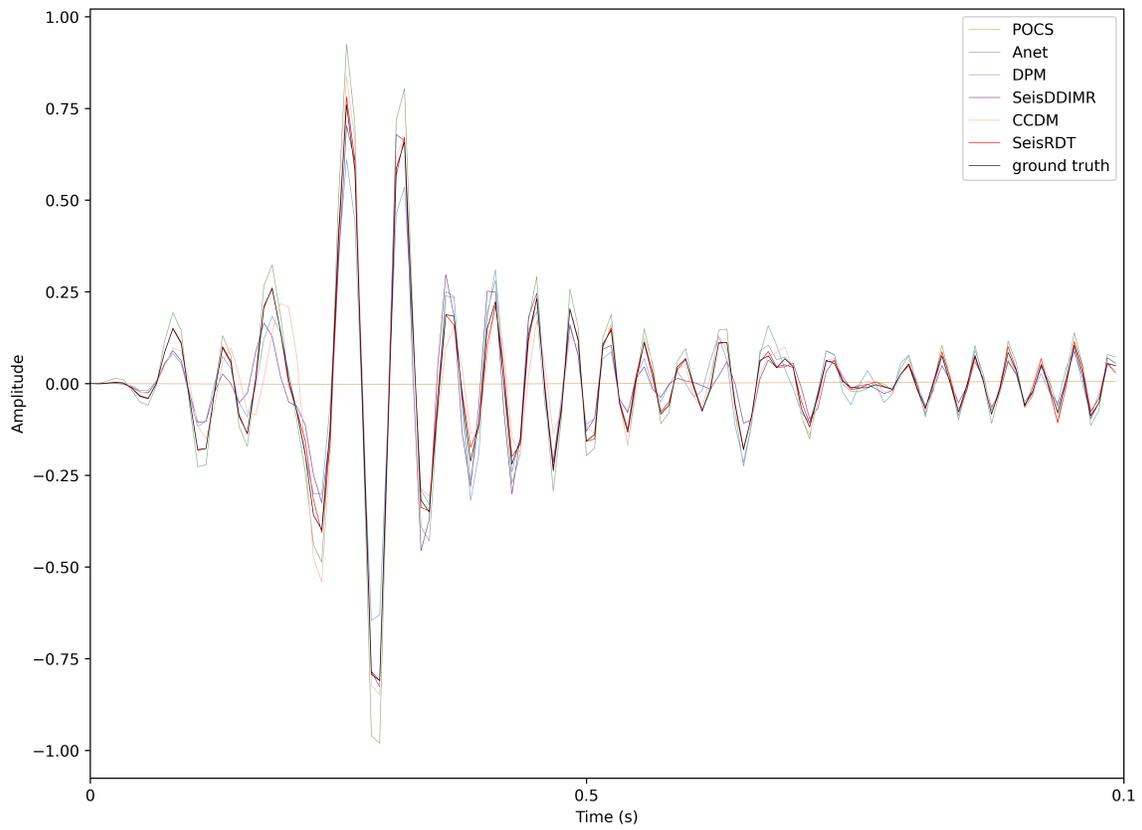
{Comparison of the 100th trace.}
\subsubsection{Field Dataset}
To thoroughly validate the effectiveness of the proposed method, we also designed two missing cases: random missing and continuous missing.

Random missing:
We set two missing proportions, 40\% and 70\%, by setting trace to 0 to simulate missing traces. Fig~\ref{fig:10} shows the reconstruction results of different methods under these two missing rates. From the residual plots, it can be observed that the deviation between the SeisRDT reconstruction results and the ground truth is the smallest, indicating the highest reconstruction accuracy.
\plot{10}{width=1.1\textwidth}
{(a) Ground truth,(b) 40\% random missing, (b-1) to (b-6) are the reconstruction results of POCS, Anet, DPM, SeisDDIMR, CCDM, and SeisRDT, respectively. (b-7) to (b-12) are the residuals of each reconstruction result relative to the ground truth.(c) 70\% random missing, (c-1) to (c-6) are the reconstruction results of POCS, Anet, DPM, SeisDDIMR, CCDM, and SeisRDT, respectively. (c-7) to (c-12) are the residuals of each reconstruction result relative to the ground truth.}

To quantitatively evaluate the quality of the reconstruction results from the six methods, we computed three evaluation metrics, as shown in Table~\ref{tbl:3}. It can be seen that our method demonstrates improvements compared to other methods at different missing rates.
\tabl{3}{Comparison of six reconstruction methods under random missing on filed dataset. The upper section compares 40\% randomly missing traces, while the lower section compares 70\% randomly missing traces.}{
	\begin{center}
		\linespread{1.25} \selectfont
		\begin{tabular}{ccccc}
			\hline
			&                 & MSE       & SNR     & SSIM   \\ \hline
			& POCS  & 8.4441e-3    & 34.9790 & 0.9438 \\
			& Anet & 4.0096e-3 & 36.4278 & 0.9633 \\   
			& DPM            & 8.8499e-5    & 46.5512 & 0.9955 \\
			& SeisDDIMR & 9.0741e-5 & 46.4425 & 0.9949 \\
			& CCDM         & 8.2921e-5 & 46.8839 & 0.9953 \\
			& SeisRDT        & 1.0948e-5 & 50.6269 & 0.9982 \\ \hline
			& POCS  & 0.0139    & 26.7548 & 0.7630 \\
			& Anet & 9.1048e-3 & 34.5613 & 0.9503 \\  
			& DPM            & 1.7335e-4    & 43.6311  & 0.9893 \\
			& SeisDDIMR & 1.8052e-4    & 43.4552 & 0.9892 \\
			& CCDM         & 1.6682e-4 & 43.7980 & 0.9898 \\
			& SeisRDT        & 4.1590e-5 & 47.0052 & 0.9919 \\ \hline
		\end{tabular}
	\end{center}
}

Continuous Missing:
We set two scenarios of continuous missing data: one with 25 missing traces and the other with 45 missing traces, where traces were set to 0 to simulate the missing traces. Fig~\ref{fig:11} illustrates the reconstruction results of different methods.
\plot{11}{width=1.1\textwidth}
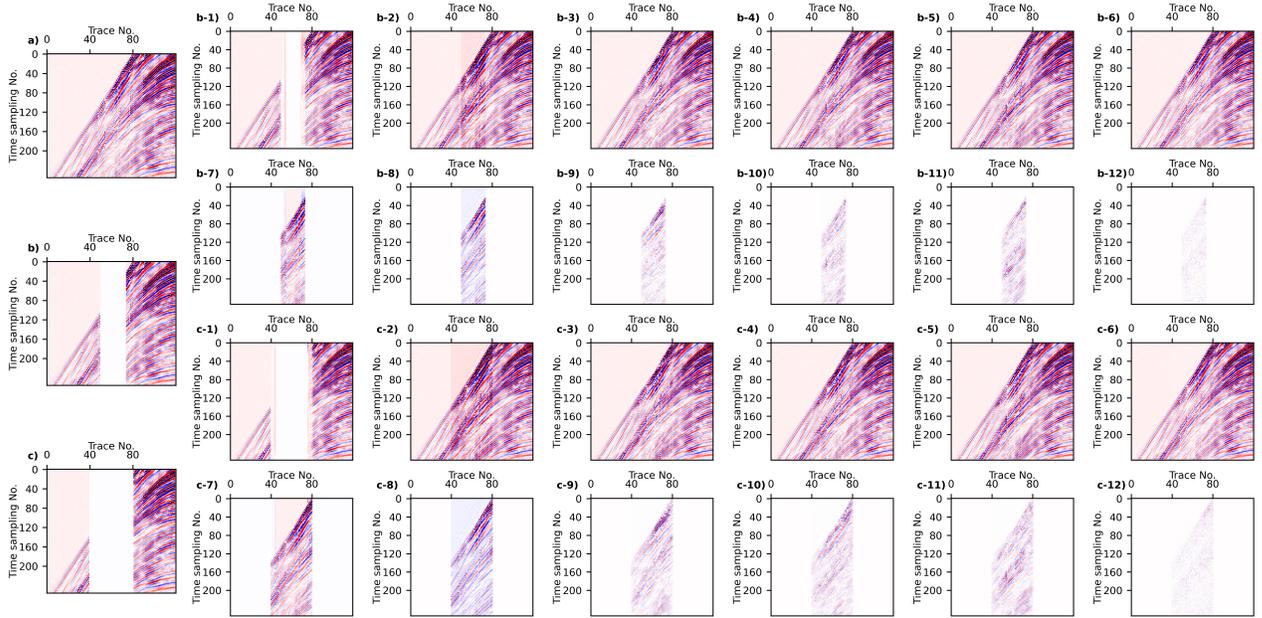
{(a) Ground truth,(b) 25 continuous missing traces, (b-1) to (b-6) are the reconstruction results of POCS, Anet, DPM, SeisDDIMR, CCDM, and SeisRDT, respectively. (b-7) to (b-12) are the residuals of each reconstruction result relative to the ground truth.(c) 45 continuous missing traces, (c-1) to (c-6) are the reconstruction results of POCS, Anet, DPM, SeisDDIMR, CCDM, and SeisRDT, respectively. (c-7) to (c-12) are the residuals of each reconstruction result relative to the ground truth.}

According to the residual plots, it can be observed that our method has the smallest bias. Table~\ref{tbl:4} presents the quantitative results of the six methods, indicating that SeisRDT yields the best reconstruction outcomes.
\tabl{4}{Comparison of six reconstruction methods under continuous missing on filed dataset. The upper section compares 25 continuous missing traces, while the lower section compares 45 continuous missing traces.}{
	\begin{center}
		\linespread{1.25} \selectfont
		\begin{tabular}{ccccc}
			\hline
			&                 & MSE       & SNR     & SSIM   \\ \hline
			& POCS  & 0.0448    & 19.5007 & 0.9092 \\
			& Anet & 8.4499e-4 & 39.7803 & 0.9629 \\   
			& DPM            & 1.5737e-4    & 44.0512 & 0.9921 \\
			& SeisDDIMR & 9.4046e-5 & 46.2871 & 0.9937 \\
			& CCDM         & 1.3870e-4 & 44.5995 & 0.9913 \\
			& SeisRDT        & 2.2664e-5 & 48.9948 & 0.9988 \\ \hline
			& POCS  & 0.1174    & 15.3212 & 0.8418 \\
			& Anet & 3.4872e-3 & 36.7502 & 0.9481 \\  
			& DPM            & 2.8592e-4    & 41.4580  & 0.9862 \\
			& SeisDDIMR & 2.1141e-4    & 42.7691 & 0.9883 \\
			& CCDM         & 2.1340e-4 & 42.7285 & 0.9876 \\
			& SeisRDT        & 4.2940e-5 & 46.4146 & 0.9957 \\ \hline
		\end{tabular}
	\end{center}
}

To better compare the six methods, we plotted the f-k spectra under the condition of 45 continuous missing traces in Fig~\ref{fig:12}. Compared with other methods, SeisRDT exhibits the best recovery of details, indicating the effectiveness of the SeisRDT in reconstructing seismic data with missing traces.
\plot{12}{width=\textwidth}
{f-k spectra. (a) 70 continuous missing traces, (b) reconstruction result of POCS, (c) reconstruction result of Anet, (d) reconstruction result of DPM, (e) reconstruction result of SeisDDIMR, (f) reconstruction result of CCDM, (g) reconstruction result of SeisRDT, (h) ground truth.}

Fig~\ref{fig:13} displays the 60th trace reconstructed by the six different methods under the condition of 45 continuous missing traces. It can be observed that the SeisRDT exhibits better reconstruction of local detailed information, and the reconstructed data closely resemble the original seismic records.
\plot{13}{width=\textwidth}
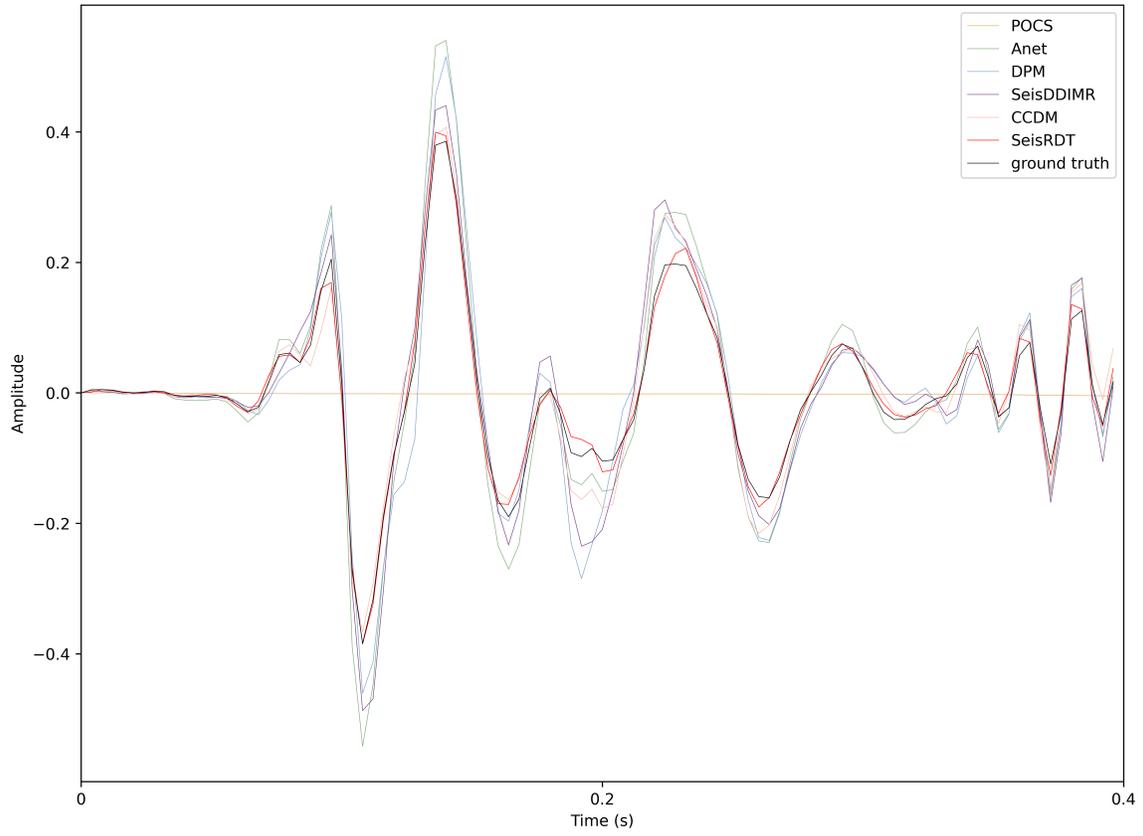
{Comparison of the 60th trace.}

The comprehensive experiments on synthetic and field datasets indicate that the diffusion model based on the UNet architecture, due to its convolutional operations' locality and the reconstruction method involving data concatenation after sampling, results in deviations between the reconstructed output and the original data. With the introduction of diffusion transformer, the diffusion model gains the capability to globally model seismic data, exhibiting stronger learning capabilities compared to the UNet architecture. Furthermore, by utilizing representation modules to infer unknown data through existing data rather than simple concatenation, the diffusion model achieves reconstructed data with a more consistent data distribution, better correlation with known data, and higher precision.
\subsection{Computational Efficiency}
To verify that SeisRDT reduces computational costs compared to other diffusion model methods, we evaluated four diffusion model methods—SeisRDT, DPM, CCDM, and SeisDDIMR—and recorded GPU memory consumption during training and inference time, we also included the computational costs of POCS and Anet to allow for a comprehensive comparison of these methods. The hardware and software configurations for both training and inference were consistent, as follows: one NVIDIA A6000 GPU and an Intel Xeon 64-core CPU. We used PyTorch 2.2.1 and CUDA 11.8 for training and inference. During training, data of size 256×256 from the SEG C3 dataset were used with batch size of 8. The GPU memory consumption during training is shown in Table~\ref{tbl:5}. As can be seen, our method consumes the least GPU memory, benefiting from the data compression model, which compresses the data into a 32×32 latent space for forward noise addition and reverse generation. CCDM consumes the most GPU memory, largely due to its constrained diffusion model being implemented with two networks—one for handling constraints and the other for data processing. In contrast, the constrained diffusion model in DPM places constraints and data into separate channels, using a two-channel input to complete the task. This results in GPU memory consumption comparable to SeisDDIMR, a model without constraints, but still higher than our method. Our implementation of the POCS method does not consume GPU memory.

We then measured the inference time for data reconstruction using the six methods. The reconstruction task involved handling 40 continuous missing traces from the SEG C3 dataset. The comparison of inference times is presented in Table~\ref{tbl:5}. As can be seen, both CCDM and SeisDDIMR have longer reconstruction times due to their iterative sampling processes, which are introduced by modifications to the sampling process. In contrast, DPM and our method have shorter inference times. Thanks to the data compression model, our method achieves a shorter reconstruction time than the other diffusion models. The Anet has the shortest reconstruction time.
\tabl{5}{Comparison of computational costs among six methods.}{
	\begin{center}
		\linespread{1.25} \selectfont
			\begin{tabular}{ccccccc}
			\hline
			           &POCS &Anet  & DPM               & SeisDDIMR   & CCDM   &  SeisRDT   \\ \hline
Reconstruction Time (s)&3.24 &0.27  & 33.81             & 154.62     & 300.75  & 18.38  \\
{\begin{tabular}[c]{@{}c@{}} GPU Memory Consumption\\ For Training (batch size 8) (GB) \end{tabular}}	
			           & 0   &7.3  & 30.6             & 30.2      & 45.3   & 23.3 \\ \hline 
		\end{tabular}
	\end{center}
}

\subsection{Continuous missing at different positions}
	To evaluate whether our method can handle continuous missing data at various positions, we set up three different continuous missing scenarios for the field dataset. In all three cases, 40 continuous traces were missing, with the missing positions set at the left, middle, and right sides, respectively. The reconstruction results using SeiRDT are shown in Fig.\ref{fig:14}. It can be observed that our method is able to effectively reconstruct the data under different missing scenarios. 
	\plot{14}{width=\textwidth}
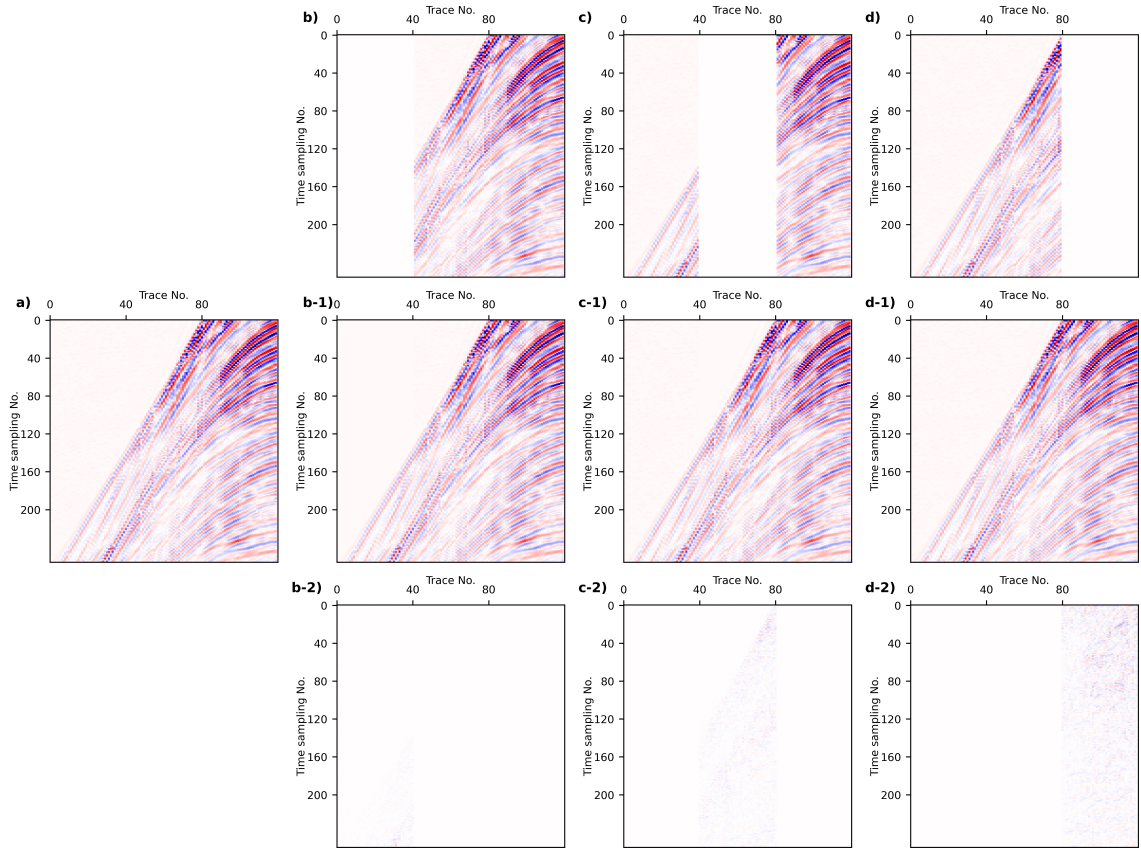
	{(a) Ground truth, (b) 40 continuous missing traces on the left, (b-1) are the reconstruction results of SeisRDT. (b-2) are the residuals of reconstruction result relative to the ground truth. (c) 40 continuous missing traces in the middle, (c-1) are the reconstruction results of SeisRDT. (c-2) are the residuals of reconstruction result relative to the ground truth. (d) 40 continuous missing traces on the right, (b-1) are the reconstruction results of SeisRDT. (b-2) are the residuals of reconstruction result relative to the ground truth.}
	
	We quantitatively calculated the results for the various missing cases, as shown in Table~\ref{tbl:6}. It can be observed that the reconstruction performance does not decline with changes in the missing data position. The slight decrease in quantitative metrics is due to the varying amounts of data to be reconstructed. The missing section on the left requires the least data to be reconstructed, while the right side requires the most. However, the overall reconstruction accuracy has not decreased. Therefore, our method is not affected by the complexity of irregular missing data.
\tabl{6}{continuous missing data scenarios at different locations in the field dataset.}{
	\begin{center}
		\linespread{1.25} \selectfont
		\begin{tabular}{ccccc}
		\hline
		&                 & MSE       & SNR     & SSIM   \\ \hline
		& Left            & 1.4598e-5    & 54.6298 & 0.9984 \\
		& Middle & 2.3412e-5 & 53.4146 & 0.9976 \\
		& Right        & 3.5532e-5 & 52.2190 & 0.9963 \\ \hline
		\end{tabular}
	\end{center}
}

\subsection{Generalization}
To investigate the generalization capability of the proposed method, we conducted generalization experiments. The training set for the experiments consisted of synthetic data from SEG C3. We then used the model trained on SEG C3 to reconstruct a more complex field dataset. The data are collected from the North China Plain in eastern China. The North China Plain is a Cenozoic faulted basin located on the North China Platform. During the Late Tertiary and Quaternary periods, a vast, contiguous plain was formed, while the mountain blocks at the edges of the plain were relatively uplifted. The Cenozoic era experienced subsidence, with thick sedimentation, and in some areas, sediment layers reached up to a kilometer in thickness. The stratigraphy develops from the bottom to the top, including the Archaean-Proterozoic, Middle to Late Neoproterozoic, Lower Paleozoic, Upper Paleozoic, Mesozoic, and Cenozoic eras. We simulated continuous missing data by removing 50 traces, setting them to 0 to represent missing data. The reconstruction results from five methods are shown in Fig. \ref{fig:15}. It can be observed that, due to the diffusion model’s ability to learn the probability distribution of the data and provide ideal coverage of the distribution, diffusion model-based methods are able to reconstruct some fundamental data structure features. SeisDDIMR shows slightly poorer generalization, likely because it only modifies the sampling process without using input data as constraints, resulting in less effective use of known data and poorer performance in low amplitude regions. Both DPM and CCDM utilize known data as constraints during reconstruction, which allows for some reconstruction of low amplitude areas. Our method, which is based entirely on known data for inference, maximizes the use of available information, showing better generalization and reconstructing the general structural features of the data. However, the precision remains poor in terms of finer details. It is evident that the slope of the data reconstructed by SeisRDT does not match the true values, while DPM and CCDM exhibit better performance in slope prediction. However, SeisRDT demonstrates greater accuracy in reconstructing low amplitude regions. Anet performs well in reconstructing high amplitude data but shows poorer results for low amplitude data.
\plot{15}{width=\textwidth}
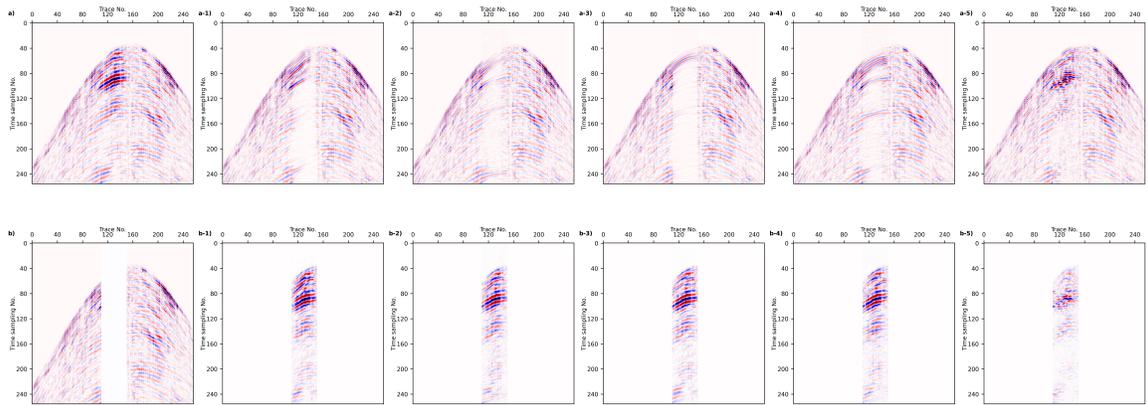
{(a) Ground truth,(b) 50 continuous missing traces, (a-1) to (a-5) are the reconstruction results of Anet, DPM, SeisDDIMR, CCDM, and SeisRDT, respectively. (b-1) to (b-5) are the residuals of each reconstruction result relative to the ground truth.}

We calculated three evaluation metrics in Table~\ref{tbl:7} to quantitatively assess the generalization capability of the five methods. The results show that our method has higher accuracy compared to the others, as the representation module uses known data exclusively for inferring and reconstructing unknown data. However, since the diffusion model reconstructs data based on its distribution, the accuracy of reconstruction may further deteriorate when dealing with data distributions that are markedly different from those of the training data.
\tabl{7}{Comparison of the generalization capabilities of five reconstruction networks with 50 continuous missing traces in the field dataset.}{
	\begin{center}
		\linespread{1.25} \selectfont
		\begin{tabular}{ccccc}
			\hline
		&                 & MSE       & SNR     & SSIM   \\ \hline
		& Anet            & 0.0014    & 34.6084 & 0.9358 \\
        & DPM            & 0.0009    & 35.1440 & 0.9448 \\
		& SeisDDIMR & 0.0016 & 33.7671 & 0.9376 \\
		& CCDM         & 0.0008 & 35.7281 & 0.9466 \\
		& SeisRDT        & 0.0005 & 37.2578 & 0.9620 \\ \hline
		\end{tabular}
	\end{center}
}

We then used the model trained on the SEG C3 dataset to reconstruct the Mobil Avo Viking Graben Line 12 field dataset. We simulated missing data by removing 25 consecutive traces and setting the missing traces to 0 as the input for reconstruction. The reconstruction results of five methods are shown in Fig. \ref{fig:16}. It can be observed that none of the five methods successfully reconstructed the data. This is because diffusion models learn the probability distribution of the data, and the probability distribution of the Mobil Avo Viking Graben Line 12 dataset is entirely different from that of SEG C3, leading to reconstruction failure. This result is consistent with the conclusions of DPM \cite[]{liu2024generative}. The significant difference in distributions causes the diffusion models' direct sampling results to contain considerable noise. In the case of SeisRDT, the reconstruction is chaotic because the representation module's output in the latent space is also highly noisy, resulting in the data decoded by the compression model being disordered and unusable. Anet also failed to reconstruct the data because it is based on convolution and employs a point-to-point reconstruction approach, learning features from the training set. When applied to unseen data with significant differences, this point-to-point reconstruction method can only mechanically reconstruct based on the features from the training set, thus failing to generalize effectively.
\plot{16}{width=\textwidth}
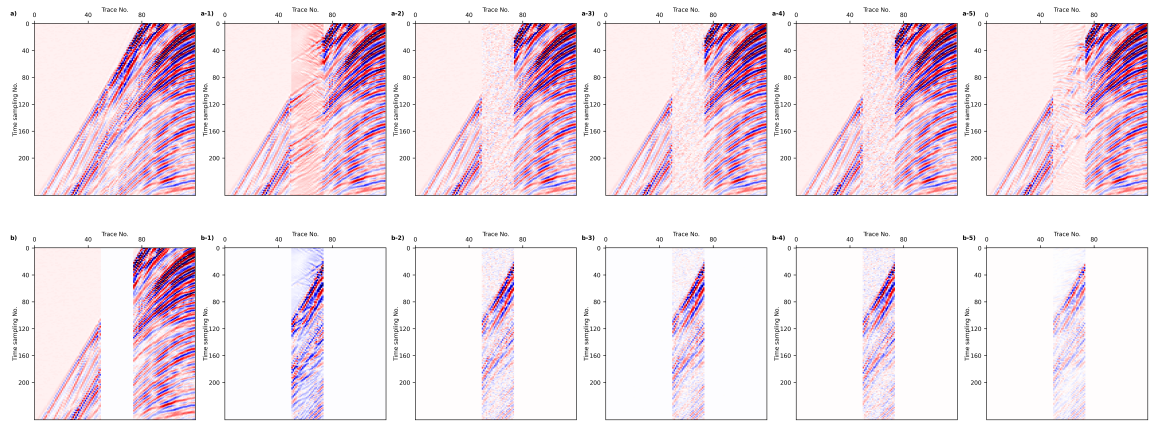
{(a) Ground truth, (b) 25 continuous missing traces, (a-1) to (a-5) are the reconstruction results of Anet, DPM, SeisDDIMR, CCDM, and SeisRDT, respectively. (b-1) to (b-5) are the residuals of each reconstruction result relative to the ground truth.}

In Table~\ref{tbl:8}, we calculated three evaluation metrics to quantitatively assess the generalization capability of the five methods. The results indicate that when the distribution of the reconstructed data is entirely different from that of the training set, diffusion models struggle to successfully reconstruct the data.
	\tabl{8}{Comparison of the generalization capabilities of five reconstruction networks with 25 continuous missing traces in the Mobil Avo Viking Graben Line 12 dataset.}{
	\begin{center}
	\linespread{1.25} \selectfont
	\begin{tabular}{ccccc}
		\hline
		&                 & MSE       & SNR     & SSIM   \\ \hline
		& Anet            & 0.0011    & 32.4441 & 0.9119 \\
		& DPM            & 0.0023    & 30.2756 & 0.8986 \\
		& SeisDDIMR  & 0.0016 & 29.3180 & 0.9071 \\
		& CCDM         & 0.0017 & 30.3457 & 0.9041 \\
		& SeisRDT        & 0.0014 & 30.5465 & 0.9043 \\ \hline
	\end{tabular}
\end{center}
}

\section{Conclusions}
This paper proposes a method for reconstructing complex and irregularly missing seismic data using representation learning-based latent diffusion transform. By compressing the original data into a latent space, compared to other reconstruction methods based on diffusion models, it significantly reduces computational and inference costs. The use of the diffusion transform architecture endows the diffusion model with the capability to globally model seismic data, exhibiting stronger learning and generalization capabilities compared to the convolutional operation-based UNet architecture. A masked modeling scheme based on representation learning is introduced, which utilizes representation modules to infer unknown data through existing data, leading to reconstructed data with a more consistent data distribution, better correlation with known data, and higher precision. Comparative experimental results on synthetic and field datasets demonstrate that our proposed method achieves more accurate interpolation results compared to other existing methods. Moreover, the diffusion model itself, by learning the data distribution, possesses higher generative accuracy and generalization capability, allowing our network to generalize to more complex missing scenarios during inference. However, because the diffusion model learns the probability distribution of the data, it cannot generalize to datasets whose distributions are entirely different from the original dataset. When the distribution of the reconstructed data differs significantly from that of the training set, performance is limited.

\verbatiminput{geophysics_example.ltx}



\newpage

\bibliographystyle{seg}  
\bibliography{example}

\begin{thebibliography}{}
\itemsep0pt

\bibitem[Cao et~al., 2022]{cao2022swin}
Cao, H., Y. Wang, J. Chen, D. Jiang, X. Zhang, Q. Tian, and M. Wang,  2022,
  Swin-unet: Unet-like pure transformer for medical image segmentation:
  European conference on computer vision, Springer, 205--218.

\bibitem[Chang et~al., 2020]{chang2020seismic}
Chang, D., W. Yang, X. Yong, G. Zhang, W. Wang, H. Li, and Y. Wang,  2020,
  Seismic data interpolation using dual-domain conditional generative
  adversarial networks: IEEE Geoscience and Remote Sensing Letters, {\bfseries
  18}, 1856--1860.

\bibitem[Chen et~al., 2021]{chen2021transunet}
Chen, J., Y. Lu, Q. Yu, X. Luo, E. Adeli, Y. Wang, L. Lu, A.~L. Yuille, and Y.
  Zhou,  2021, Transunet: Transformers make strong encoders for medical image
  segmentation: arXiv preprint arXiv:2102.04306.

\bibitem[Chen et~al., 2019]{chen2019interpolation}
Chen, Y., X. Chen, Y. Wang, and S. Zu,  2019, The interpolation of sparse
  geophysical data: Surveys in Geophysics, {\bfseries 40}, 73--105.

\bibitem[Cheng et~al., 2021]{cheng2021task}
Cheng, G., R. Li, C. Lang, and J. Han,  2021, Task-wise attention guided part
  complementary learning for few-shot image classification: Science China
  Information Sciences, {\bfseries 64}, 120104.

\bibitem[Creswell et~al., 2018]{creswell2018generative}
Creswell, A., T. White, V. Dumoulin, K. Arulkumaran, B. Sengupta, and A.~A.
  Bharath,  2018, Generative adversarial networks: An overview: IEEE signal
  processing magazine, {\bfseries 35}, 53--65.

\bibitem[Deng et~al., 2024]{deng2024seismic}
Deng, F., S. Wang, X. Wang, and P. Fang,  2024, Seismic data reconstruction
  based on conditional constraint diffusion model: IEEE Geoscience and Remote
  Sensing Letters.

\bibitem[Devlin et~al., 2018]{devlin2018bert}
Devlin, J., M.-W. Chang, K. Lee, and K. Toutanova,  2018, Bert: Pre-training of
  deep bidirectional transformers for language understanding: arXiv preprint
  arXiv:1810.04805.

\bibitem[Dhariwal and Nichol, 2021]{dhariwal2021diffusion}
Dhariwal, P., and A. Nichol,  2021, Diffusion models beat gans on image
  synthesis: Advances in neural information processing systems, {\bfseries 34},
  8780--8794.

\bibitem[Dosovitskiy et~al., 2020]{dosovitskiy2020image}
Dosovitskiy, A., L. Beyer, A. Kolesnikov, D. Weissenborn, X. Zhai, T.
  Unterthiner, M. Dehghani, M. Minderer, G. Heigold, S. Gelly, et~al.,  2020,
  An image is worth 16x16 words: Transformers for image recognition at scale:
  arXiv preprint arXiv:2010.11929.

\bibitem[Esser et~al., 2021]{esser2021taming}
Esser, P., R. Rombach, and B. Ommer,  2021, Taming transformers for
  high-resolution image synthesis: Proceedings of the IEEE/CVF conference on
  computer vision and pattern recognition, 12873--12883.

\bibitem[Fomel, 2003]{fomel2003seismic}
Fomel, S.,  2003, Seismic reflection data interpolation with differential
  offset and shot continuation: Geophysics, {\bfseries 68}, 733--744.

\bibitem[Gao et~al., 2023]{gao2023masked}
Gao, S., P. Zhou, M.-M. Cheng, and S. Yan,  2023, Masked diffusion transformer
  is a strong image synthesizer: Proceedings of the IEEE/CVF International
  Conference on Computer Vision, 23164--23173.

\bibitem[G{\"u}l{\"u}nay, 2003]{gulunay2003seismic}
G{\"u}l{\"u}nay, N.,  2003, Seismic trace interpolation in the fourier
  transform domain: Geophysics, {\bfseries 68}, 355--369.

\bibitem[He et~al., 2022]{he2022masked}
He, K., X. Chen, S. Xie, Y. Li, P. Doll{\'a}r, and R. Girshick,  2022, Masked
  autoencoders are scalable vision learners: Proceedings of the IEEE/CVF
  conference on computer vision and pattern recognition, 16000--16009.

\bibitem[He et~al., 2016]{he2016deep}
He, K., X. Zhang, S. Ren, and J. Sun,  2016, Deep residual learning for image
  recognition: Proceedings of the IEEE conference on computer vision and
  pattern recognition, 770--778.

\bibitem[He et~al., 2021]{he2021seismic}
He, T., B. Wu, and X. Zhu,  2021, Seismic data consecutively missing trace
  interpolation based on multistage neural network training process: IEEE
  Geoscience and Remote Sensing Letters, {\bfseries 19}, 1--5.

\bibitem[Ho et~al., 2020]{ho2020denoising}
Ho, J., A. Jain, and P. Abbeel,  2020, Denoising diffusion probabilistic
  models: Advances in Neural Information Processing Systems, {\bfseries 33},
  6840--6851.

\bibitem[Huang et~al., 2022]{9703292}
Huang, H., T. Wang, J. Cheng, Y. Xiong, C. Wang, and J. Geng,  2022,
  Self-supervised deep learning to reconstruct seismic data with consecutively
  missing traces: IEEE Transactions on Geoscience and Remote Sensing,
  {\bfseries 60}, 1--14.

\bibitem[Innocent~Obou{\'e} et~al., 2021]{innocent2021robust}
Innocent~Obou{\'e}, Y. A.~S., W. Chen, H. Wang, and Y. Chen,  2021, Robust
  damped rank-reduction method for simultaneous denoising and reconstruction of
  5d seismic data: Geophysics, {\bfseries 86}, V71--V89.

\bibitem[Jia and Ma, 2017]{jia2017can}
Jia, Y., and J. Ma,  2017, What can machine learning do for seismic data
  processing? an interpolation application: Geophysics, {\bfseries 82},
  V163--V177.

\bibitem[Kaur et~al., 2021]{kaur2021seismic}
Kaur, H., N. Pham, and S. Fomel,  2021, Seismic data interpolation using deep
  learning with generative adversarial networks: Geophysical Prospecting,
  {\bfseries 69}, 307--326.

\bibitem[Kingma et~al., 2021]{kingma2021variational}
Kingma, D., T. Salimans, B. Poole, and J. Ho,  2021, Variational diffusion
  models: Advances in neural information processing systems, {\bfseries 34},
  21696--21707.

\bibitem[Latif and Mousa, 2016]{latif2016efficient}
Latif, A., and W.~A. Mousa,  2016, An efficient undersampled high-resolution
  radon transform for exploration seismic data processing: IEEE Transactions on
  Geoscience and Remote Sensing, {\bfseries 55}, 1010--1024.

\bibitem[Li et~al., 2021]{li2021survey}
Li, Z., F. Liu, W. Yang, S. Peng, and J. Zhou,  2021, A survey of convolutional
  neural networks: analysis, applications, and prospects: IEEE transactions on
  neural networks and learning systems, {\bfseries 33}, 6999--7019.

\bibitem[Liu and Ma, 2024]{liu2024generative}
Liu, Q., and J. Ma,  2024, Generative interpolation via a diffusion
  probabilistic model: Geophysics, {\bfseries 89}, V65--V85.

\bibitem[Liu et~al., 2021]{liu2021swin}
Liu, Z., Y. Lin, Y. Cao, H. Hu, Y. Wei, Z. Zhang, S. Lin, and B. Guo,  2021,
  Swin transformer: Hierarchical vision transformer using shifted windows:
  Proceedings of the IEEE/CVF international conference on computer vision,
  10012--10022.

\bibitem[Ma et~al., 2024]{ma2024sit}
Ma, N., M. Goldstein, M.~S. Albergo, N.~M. Boffi, E. Vanden-Eijnden, and S.
  Xie,  2024, Sit: Exploring flow and diffusion-based generative models with
  scalable interpolant transformers: arXiv preprint arXiv:2401.08740.

\bibitem[Mirza and Osindero, 2014]{mirza2014conditional}
Mirza, M., and S. Osindero,  2014, Conditional generative adversarial nets:
  arXiv preprint arXiv:1411.1784.

\bibitem[Mousavi and Beroza, 2022]{mousavi2022deep}
Mousavi, S.~M., and G.~C. Beroza,  2022, Deep-learning seismology: Science,
  {\bfseries 377}, eabm4470.

\bibitem[Mousavi et~al., 2020]{mousavi2020earthquake}
Mousavi, S.~M., W.~L. Ellsworth, W. Zhu, L.~Y. Chuang, and G.~C. Beroza,  2020,
  Earthquake transformer—an attentive deep-learning model for simultaneous
  earthquake detection and phase picking: Nature communications, {\bfseries
  11}, 3952.

\bibitem[Niu et~al., 2021]{niu2021seismic}
Niu, X., L. Fu, W. Zhang, and Y. Li,  2021, Seismic data interpolation based on
  simultaneously sparse and low-rank matrix recovery: IEEE Transactions on
  Geoscience and Remote Sensing, {\bfseries 60}, 1--13.

\bibitem[Peebles and Xie, 2023]{peebles2023scalable}
Peebles, W., and S. Xie,  2023, Scalable diffusion models with transformers:
  Proceedings of the IEEE/CVF International Conference on Computer Vision,
  4195--4205.

\bibitem[Porsani, 1999]{porsani1999seismic}
Porsani, M.~J.,  1999, Seismic trace interpolation using half-step prediction
  filters: Geophysics, {\bfseries 64}, 1461--1467.

\bibitem[Radford et~al., 2018]{radford2018improving}
Radford, A., K. Narasimhan, T. Salimans, I. Sutskever, et~al.,  2018, Improving
  language understanding by generative pre-training.

\bibitem[Rombach et~al., 2022]{rombach2022high}
Rombach, R., A. Blattmann, D. Lorenz, P. Esser, and B. Ommer,  2022,
  High-resolution image synthesis with latent diffusion models: Proceedings of
  the IEEE/CVF conference on computer vision and pattern recognition,
  10684--10695.

\bibitem[Ronen, 1987]{ronen1987wave}
Ronen, J.,  1987, Wave-equation trace interpolation: Geophysics, {\bfseries
  52}, 973--984.

\bibitem[Ronneberger et~al., 2015]{ronneberger2015u}
Ronneberger, O., P. Fischer, and T. Brox,  2015, U-net: Convolutional networks
  for biomedical image segmentation: Medical image computing and
  computer-assisted intervention--MICCAI 2015: 18th international conference,
  Munich, Germany, October 5-9, 2015, proceedings, part III 18, Springer,
  234--241.

\bibitem[Sohl-Dickstein et~al., 2015]{sohl2015deep}
Sohl-Dickstein, J., E. Weiss, N. Maheswaranathan, and S. Ganguli,  2015, Deep
  unsupervised learning using nonequilibrium thermodynamics: International
  conference on machine learning, PMLR, 2256--2265.

\bibitem[Song et~al., 2020]{song2020denoising}
Song, J., C. Meng, and S. Ermon,  2020, Denoising diffusion implicit models:
  arXiv preprint arXiv:2010.02502.

\bibitem[Song and Ermon, 2019]{song2019generative}
Song, Y., and S. Ermon,  2019, Generative modeling by estimating gradients of
  the data distribution: Advances in neural information processing systems,
  {\bfseries 32}.

\bibitem[Vahdat et~al., 2021]{vahdat2021score}
Vahdat, A., K. Kreis, and J. Kautz,  2021, Score-based generative modeling in
  latent space: Advances in neural information processing systems, {\bfseries
  34}, 11287--11302.

\bibitem[Wang et~al., 2022]{9768816}
Wang, B., D. Han, and J. Li,  2022, Missing shots and near-offset
  reconstruction of marine seismic data with towered streamers via
  self-supervised deep learning: IEEE Transactions on Geoscience and Remote
  Sensing, {\bfseries 60}, 1--9.

\bibitem[Wang et~al., 2019]{wang2019deep}
Wang, B., N. Zhang, W. Lu, and J. Wang,  2019, Deep-learning-based seismic data
  interpolation: A preliminary result: Geophysics, {\bfseries 84}, V11--V20.

\bibitem[Wang et~al., 2010]{wang2010seismic}
Wang, J., M. Ng, and M. Perz,  2010, Seismic data interpolation by greedy local
  radon transform: Geophysics, {\bfseries 75}, WB225--WB234.

\bibitem[Wang et~al., 2024a]{wang2024seisfusion}
Wang, S., F. Deng, P. Jiang, Z. Gong, X. Wei, and Y. Wang,  2024a, Seisfusion:
  Constrained diffusion model with input guidance for 3d seismic data
  interpolation and reconstruction: IEEE Transactions on Geoscience and Remote
  Sensing.

\bibitem[Wang et~al., 2024b]{wang2024reconstructing}
Wang, X., Z. Wang, Z. Xiong, Y. Yang, C. Zhu, and J. Gao,  2024b,
  Reconstructing regularly missing seismic traces with a classifier-guided
  diffusion model: IEEE Transactions on Geoscience and Remote Sensing,
  {\bfseries 62}, 1--14.

\bibitem[Wang et~al., 2020]{wang2020seismic}
Wang, Y., B. Wang, N. Tu, and J. Geng,  2020, Seismic trace interpolation for
  irregularly spatial sampled data using convolutional autoencodercae-based
  seismic trace interpolation: Geophysics, {\bfseries 85}, V119--V130.

\bibitem[Wang et~al., 2024c]{wang2024patch}
Wang, Z., Y. Jiang, H. Zheng, P. Wang, P. He, Z. Wang, W. Chen, M. Zhou,
  et~al.,  2024c, Patch diffusion: Faster and more data-efficient training of
  diffusion models: Advances in Neural Information Processing Systems,
  {\bfseries 36}.

\bibitem[Wei et~al., 2023]{wei2023seismic}
Wei, X., C. Zhang, H. Wang, C. Tan, D. Xiong, B. Jiang, J. Zhang, and S.-W.
  Kim,  2023, Seismic data interpolation based on denoising diffusion implicit
  models with resampling: arXiv preprint arXiv:2307.04226.

\bibitem[Yang et~al., 2023]{yang2023diffusion}
Yang, L., Z. Zhang, Y. Song, S. Hong, R. Xu, Y. Zhao, W. Zhang, B. Cui, and
  M.-H. Yang,  2023, Diffusion models: A comprehensive survey of methods and
  applications: ACM Computing Surveys, {\bfseries 56}, 1--39.

\bibitem[Yu and Wu, 2021]{yu2021attention}
Yu, J., and B. Wu,  2021, Attention and hybrid loss guided deep learning for
  consecutively missing seismic data reconstruction: IEEE Transactions on
  Geoscience and Remote Sensing, {\bfseries 60}, 1--8.

\bibitem[Zhang et~al., 2020]{zhang2020can}
Zhang, H., X. Yang, and J. Ma,  2020, Can learning from natural image denoising
  be used for seismic data interpolation?: Geophysics, {\bfseries 85},
  WA115--WA136.

\bibitem[Zhang et~al., 2019]{zhang2019nonconvex}
Zhang, W., L. Fu, and Q. Liu,  2019, Nonconvex log-sum function-based
  majorization--minimization framework for seismic data reconstruction: IEEE
  Geoscience and Remote Sensing Letters, {\bfseries 16}, 1776--1780.

\end{thebibliography}

\end{document}